\documentclass[12pt]{iopart}
\usepackage{graphics}
\begin{document}
\jl{1}
\title
[Ground state of an $S=1/2$ distorted diamond chain]
{Ground state of an $S=1/2$ distorted diamond chain \\
--- model of $\rm Cu_3 Cl_6 (H_2 O)_2 \cdot 2H_8 C_4 SO_2$}
\author{Kiyomi Okamoto\dag, Taskashi Tonegawa\ddag, Yutaka Takahashi\S\  
        and Makoto Kaburagi$\|$}

\address{\dag Department of Physics,
         Tokyo Institute of Technology,
         Oh-Okayama, Meguro-ku, Tokyo 152-8551, Japan}

\address{\ddag Department of Physics, Faculty of Science, Kobe University, Rokkodai,
  Kobe 657-8501, Japan}

\address{\S Division of Physics, Graduate School of Science and Technology, Kobe
  University, Rokkodai, Kobe 657-8501, Japan}

\address{$\|$ Department of Informatics, Faculty of Cross-Cultural Studies, Kobe
  University, Tsurukabuto, Kobe 657-8501, Japan}

\begin{abstract}
We study the ground state of the model Hamiltonian of
the trimerized $S=1/2$ quantum Heisenberg chain
$\rm Cu_3 Cl_6 (H_2 O)_2 \cdot 2H_8 C_4 SO_2$ in which the non-magnetic ground
state is observed recently.
This model consists of stacked trimers and has three kinds of coupling constants
between spins;
the intra-trimer coupling constant $J_1$ and the inter-trimer coupling constants
$J_2$ and $J_3$.
All of these constants are assumed to be antiferromagnetic.
By use of the analytical method and physical considerations,
we show that there are three phases on the $\tilde J_2 - \tilde J_3$ plane
($\tilde J_2 \equiv J_2/J_1$, $\tilde J_3 \equiv J_3/J_1$),
the dimer phase, the spin fluid phase and the ferrimagnetic phase.
The dimer phase is caused by the frustration effect.
In the dimer phase, there exists the excitation gap between
the two-fold degenerate ground state and the first excited state,
which explains the non-magnetic ground state observed in 
$\rm Cu_3 Cl_6 (H_2 O)_2 \cdot 2H_8 C_4 SO_2$.
We also obtain the phase diagram on the $ \tilde J_2 - \tilde J_3$
plane from the numerical diagonalization data for finite systems by use of the
Lanczos algorithm.
\end{abstract}
\pacs{75.10.Jm, 75.40.Cx, 75.50.Ee, 75.50.Gg}
\maketitle
\section{Introduction}
In recent years low-dimensional quantum spin systems have attracted a great
deal of attention.
Very recently, Ishii et al.\cite{Ishii} have experimentally studied a trimerized
quantum spin chain $\rm Cu_3 Cl_6 (H_2 O)_2 \cdot 2H_8 C_4 SO_2$.
They have measured the temperature dependence of the spin susceptibility,
and the magnetization curve at low temperatures.
Their results show that
the magnetic susceptibility $\chi$ behaves as $\chi \to 0$ at $T \to 0$
and there exists the critical magnetic field where the magnetization
rises up from zero.
Thus they have concluded
that the ground state of this substance is nonmagnetic.
From the result of structure-analysis experiment \cite{Swank},
the chain is composed of stacked $S=1/2$ $\rm Cu^{2+}$ trimers
and separated from other chains by large molecules $\rm H_8 C_4 SO_2$.
Therefore this substance is thought to be well modeled by independent chains of
stacked trimers and they have proposed a model shown in figure 1.
\cite{Ishii}.

\begin{figure}[h]
   \begin{center}
      \scalebox{0.5}[0.5]{\includegraphics{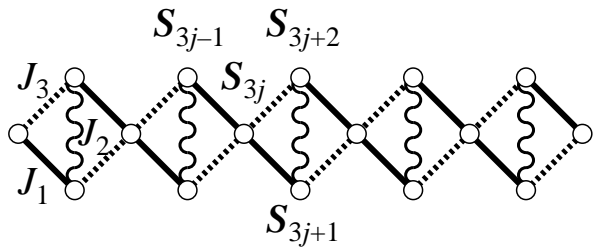}}
   \end{center}
   \caption{Sketch of the model of 
            $\rm Cu_3 Cl_6 (H_2 O)_2 \cdot 2H_8 C_4 SO_2$.
            Solid lines denote the intra-trimer coupling $J_1$,
            wavy lines the inter-trimer coupling $J_2$,
            and dotted lines the inter-trimer coupling $J_3$.}
\end{figure}

In this paper we call this model the \lq\lq distorted diamond (DD) chain model\rq\rq.
The Hamiltonian of this model is written as
\begin{eqnarray}
   H
   =  J_1 \sum_j \left( {\bi S}_{3j-1} \cdot {\bi S}_{3j} 
          + {\bi S}_{3j} \cdot {\bi S}_{3j+1} \right)
    + J_2 \sum_j {\bi S}_{3j+1} \cdot {\bi S}_{3j+2} \nonumber \\
    + J_3 \sum_j \left( {\bi S}_{3j-2} \cdot {\bi S}_{3j}
          + {\bi S}_{3j} \cdot {\bi S}_{3j+2} \right)
   \label{eq:ham}
\end{eqnarray}
where three spins ${\bi S}_{3j-1}$, ${\bi S}_{3j}$ and ${\bi S}_{3j+1}$ form a trimer.
All the coupling constants are supposed to be positive (antiferromagnetic).
Although it is thought that
$J_1 > J_2,J_3$ in $\rm Cu_3 Cl_6 (H_2 O)_2 \cdot 2H_8 C_4 SO_2$
because of its structure, we do not restrict ourselves to this case.
We note that the point $(\tilde J_2, \tilde J_3)$ is equivalent to
the point $(\tilde J_2/\tilde J_3, 1/\tilde J_3)$
by interchanging the role of $J_1$ and $J_3$.
Hereafter we take $J_1$ as the energy unit and set
$\tilde J_2 \equiv J_2/J_1$ and $\tilde J_3 \equiv J_3/J_1$.

If we transform the Hamiltonian (\ref{eq:ham}) into the fermion representation
through the Jordan-Wigner transformation, we can see that the fermionic band gap exists
at $M=M_{\rm s}/3$ but not at $M=0$, where $M_{\rm s}$ is the saturation
magnetization \cite{trimer, Okamoto-F-F-AF}.
Thus the trimerization itself cannot be the direct reason
for the nonmagnetic ground state.
This can also be explained by considering the necessary condition for the
appearance of magnetization plateau proposed by Oshikawa, Yamanaka
and Affleck \cite{OYA},
\begin{equation}
  n (S - \langle m \rangle) = {\rm integer}
  \label{eq:OYA}
\end{equation}
where $n$ is the periodicity of the ground-state wave function,
$S$ the magnitude of spins
and $\langle m \rangle$ the average magnetization per one spin in the plateau.
The periodicity of the Hamiltonian (\ref{eq:ham}) itself is $3$.
We see that $n=3$ does not satisfy the condition (\ref{eq:OYA}) with $S=1/2$ and
$\langle m \rangle =0$.
Then, if the present model is applicable to
$\rm Cu_3 Cl_6 (H_2 O)_2 \cdot 2H_8 C_4 SO_2$,
its ground-state wave function should have the periodicity
at least $n=6$ due to the spontaneous symmetry breaking.
In this paper, we explain why the non-magnetic ground state is realized
and draw the ground-state phase diagram on the $\tilde J_2 - \tilde J_3$ plane.

The $J_1 = J_3$ case of the present model was named the \lq\lq diamond chain\rq\rq
and investigated by Takano, Kubo and Sakamoto (TKS) \cite{Takano}.
They concluded that the ground state of the diamond chain is composed of
three phases;
the ferrimagnetic phase ($M=M_{\rm s}/3$) for $\tilde J_2 < 0.909$,
the tetramer-dimer phase for $0.909 < \tilde J_2 < 2$ and
the dimer-monomer phase for $\tilde J_2 > 2$.
The relation between the present model and TKS's model will be discussed later.

This paper is organized as follows.
In \S2, we explain the mechanism for the non-magnetic ground state
by use of the analytical method and a physical consideration.
In \S3, we obtain the phase diagram from the numerical data
of diagonalization of the Hamiltonian (\ref{eq:ham}) for finite systems
by use of the Lanczos algorithm.
The last section is devoted to discussions.

\section{Analytical and physical approach}

We consider three special cases at first.
When $\tilde J_2 =1$ and $\tilde J_3 =0$, the present model is reduced to the
simple $S=1/2$ chain with nearest-neighbor interactions, the ground state
of which is the spin-fluid (SF) state, as is well known.
In case of $\tilde J_2=0$, the ground state may be ferrimagnetic
($M = M_{\rm s}/3$), because the state with ${\bi S}_{3j}=\downarrow$ and
${\bi S}_{3j \pm 1} = \uparrow$ is the classical ground state.
At the point $\tilde J_2 = \tilde J_3 =0$, the chain is truncated into an
array of independent trimers.

\begin{figure}[h]
   \begin{center}
      \scalebox{0.5}[0.5]{\includegraphics{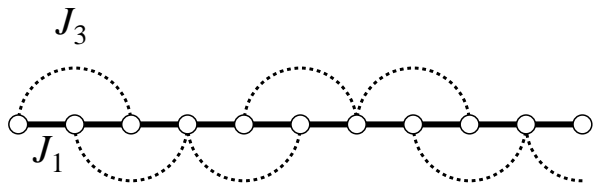}}
   \end{center}
   \caption{Present model in the $\tilde J_2=1$ case in the linear chain form.}
\end{figure}
\begin{figure}[h]
   \begin{center}
      \scalebox{0.5}[0.5]{\includegraphics{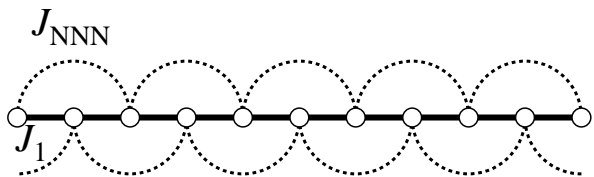}}
   \end{center}
   \caption{Next-nearest-neighbor (NNN) interaction model.}
\end{figure}

Next, let us consider the $\tilde J_2=1$ case.
If we re-draw the model in the single chain form as in figure 2,
we see this is closely related to the next-nearest-neighbor (NNN)
interaction model in figure 3.
In fact, the model of figure 2 is obtained from that of figure 3 by removing
one NNN interaction of every three NNN interactions.
The important point is that every spin feels the frustration.

The NNN interaction model is one of the most important models
having the frustration and is extensively studied
\cite{MG,Majumdar,Haldane,TH,KF,ON,THK,NO1,NO2}.
When $\tilde J_{\rm NNN} \equiv J_{\rm NNN}/J_1 = 0.5$,
the ground state of the NNN interaction model is
an array of independent singlet dimers,
where the translational symmetry by one spin spacing is spontaneously broken
\cite{MG,Majumdar}.
Then there should exist the critical point of the ground-state phase transition
between the spin-fluid (SF) state and the dimer state.
Okamoto and Nomura \cite{ON} numerically determined this SF-dimer critical point,
$\tilde J_{\rm NNN}^{\rm (cr)}=0.2411$.

Since the model of figure 2 is very similar to that of figure 3, as stated,
the SF-dimer transition may occur also in the model of figure 2 when $\tilde J_2$
is increased.
This can be confirmed by the bosonization technique in the following way.
The effective Hamilton of the model of figure 3 in the continuum limit
is written as
\cite{Haldane,KF,ON,NO1,NO2}
\begin{equation}
  H
  = \frac{1}{2\pi} \int {\rm d} x\left[ v_{\rm s}K(\pi\Pi)^{2}
    + \frac{v_{\rm s}}{K}
   \left(\frac{\partial \phi}{\partial x}\right)^{2}\right]
   + \frac{y_{\phi}v_{\rm s}}{2\pi}\int {\rm d} x\cos\sqrt{2}\phi
\label{eq:sg-nnn}
\end{equation}
where $v_{\rm s}$ is the spin wave velocity and $K$ the quantum parameter which
governs the algebraic decay of the spin correlation functions
\begin{equation}
    \langle S_0^z S_r^z \rangle \sim r^{-K},~~~~~~
    \langle S_0^+ S_r^- \rangle \sim r^{-1/K}
\end{equation}
in the SF state.
Due to the isotropic nature of our model,
the renormalized value of $K$ should be $K=1$.
The variables $\phi(x)$ and $\Pi(x)$ are mutually conjugate,
\begin{equation}
    [\phi(x),\Pi(x')] = {\rm i}\delta(x-x')
\end{equation}
The coefficient of the $\cos$-term, $y_\phi$ in equation (3), is
\begin{equation}
    y_\phi \propto \Delta - 3 \tilde J_{\rm NNN}
    \label{eq:y-nnn}
\end{equation}
where $\Delta$ is the $XXZ$ anisotropy defined by $J^z/J^\perp$ which is equal to unity
in our isotropic model.
For the model of figure 2, we can obtain the effective Hamiltonian
of the same form as (\ref{eq:sg-nnn}), but with
\begin{equation}
    y_\phi \propto \Delta - 2 \tilde J_3.
    \label{eq:y-dd}
\end{equation}
We note that the expressions (\ref{eq:y-nnn}) and (\ref{eq:y-dd}) are valid
only in the lowest order of $\Delta$, $\tilde J_{\rm NNN}$ and $\tilde J_3$.
Since we take the continuum limit in the course of deriving the effective
Hamiltonian, the difference between two models appears as the difference
in the expression of $y_\phi$.
We note that the spin wave velocity $v_{\rm s}$ is slightly different
between two models, but it does not bring about any essential effect.
Thus we can conclude that the model of figure 2 also shows the SF-dimer phase
transition.
Since $3\tilde J_{\rm NNN}$ in equation (\ref{eq:y-nnn}) is replaced by
$2\tilde J_2$ in equation (\ref{eq:y-dd}), the critical value $\tilde J_2^{\rm (cr)}$
is naively obtained by letting
$3 \tilde J_{\rm NNN}^{\rm (cr)} = 2\tilde J_2^{\rm (cr)}$,
which leads to $\tilde J_2^{\rm (cr)} \simeq 0.36$ by using
$\tilde J_{\rm NNN}^{\rm (cr)} = 0.2411$.
In fact, as shown in \S3, the numerical result is
$\tilde J_2^{\rm (cr)} \simeq 0.354$.
The dimer ground-state wave function in the model of figure 2 is two-fold degenerate
with periodicity $n=6$ and is shown in figure 4.

\begin{figure}[h]
   \begin{center}
      \scalebox{0.5}[0.5]{\includegraphics{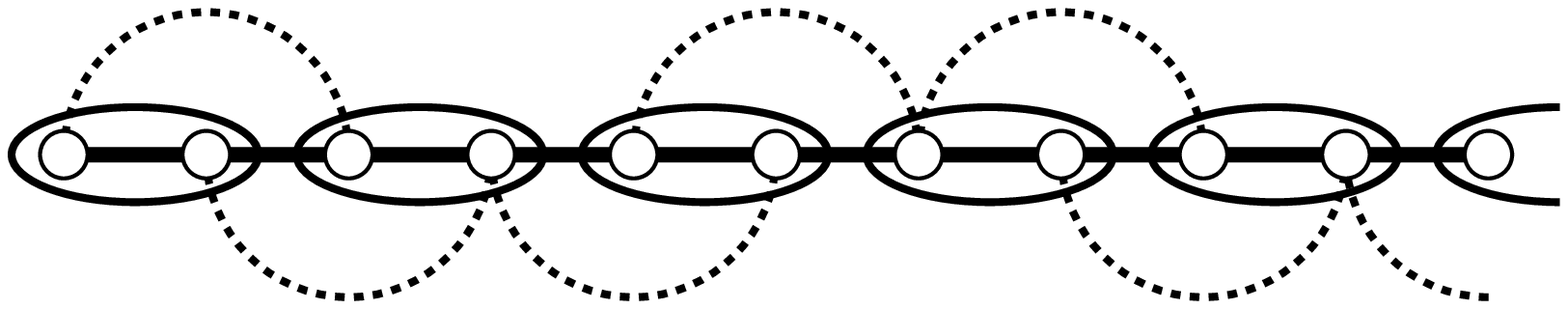}}
      \scalebox{0.5}[0.5]{\includegraphics{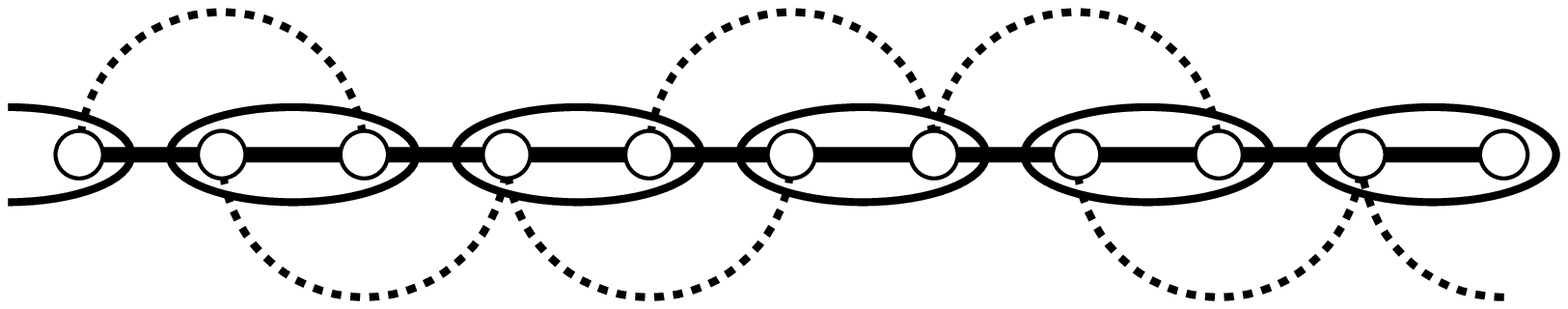}}
   \end{center}
   \caption{Dimer configurations in the ground state of the DD chain model
            with $\tilde J_2=1$ in the single chain form.
            Two spins in an ellipse form a singlet dimer pair}
\end{figure}

When $\tilde J_2 \ne 1$, we have to take the trimerization effects
into the effective Hamiltonian.
However, in the $M=0$ subspace,
the trimerization does not bring about the mass-generating term
such as $\cos$-term in equation (\ref{eq:sg-nnn}), 
although it slightly modifies the spin wave velocity $v_{\rm s}$.
Then, as far as the trimerization is not so large
(not so far from the $\tilde J_2=1$ line),
the DD chain model also exhibits the SF-dimer phase transition.
The dimer configuration of the DD chain model is easily known by tracing back
of the model mapping, which is shown in figure 5.
This ground-state wave function is also two-fold degenerate
with periodicity $n=6$.

\begin{figure}[h]
   \begin{center}
      \scalebox{0.5}[0.5]{\includegraphics{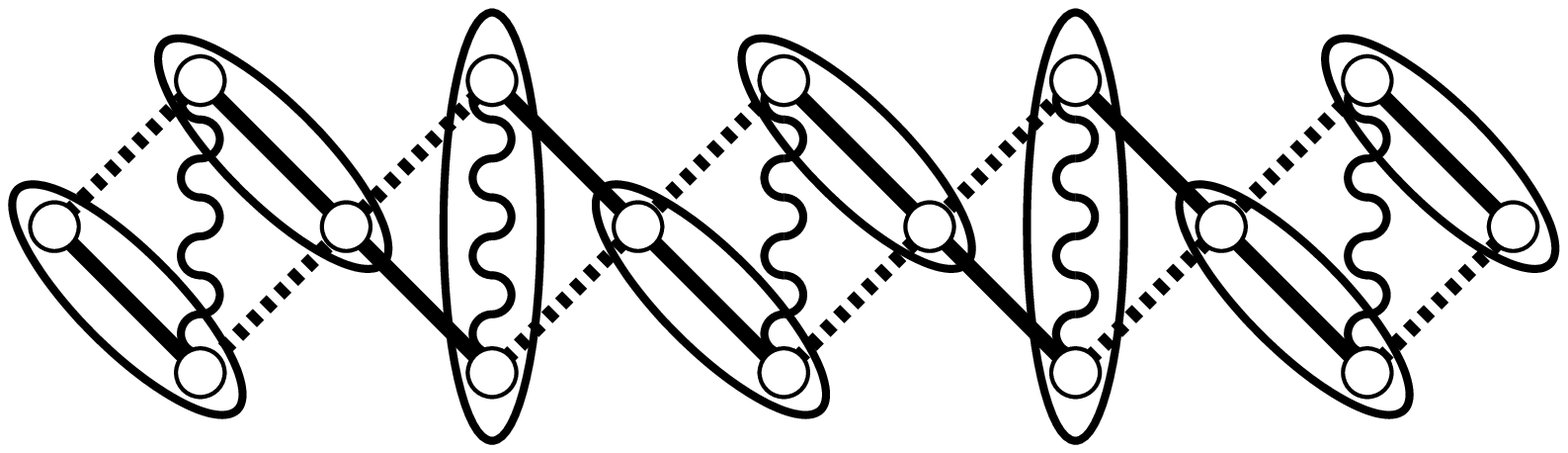}}
      \scalebox{0.5}[0.5]{\includegraphics{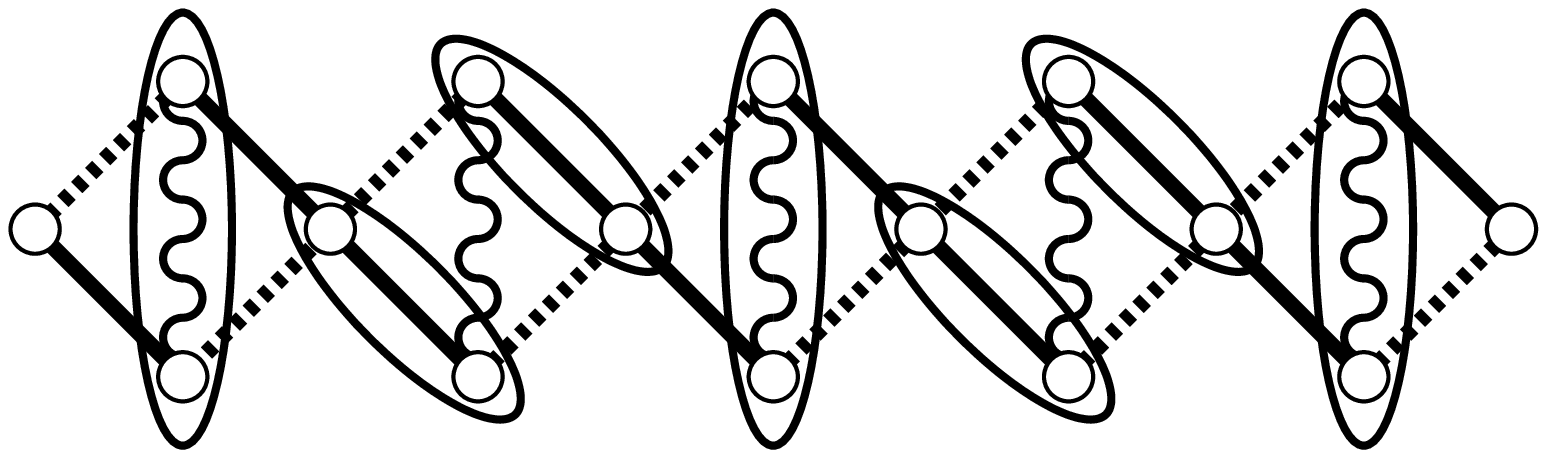}}
   \end{center}
   \caption{Dimer configurations in the ground state of the DD chain model.
            Two spins in an ellipse form a singlet dimer pair.}
\end{figure}

Here we summarize the critical properties of the SF-dimer transition
using the effective Hamiltonian (\ref{eq:sg-nnn}) having the sine-Gordon form.
The renormalization group calculation leads to
\begin{equation}
  \frac{{\rm d} y_{0}(L)}{{\rm d}\ln L} = -y_{\phi}(L)^{2}~~~~~~~~~~
  \frac{{\rm d} y_{\phi}(L)}{{\rm d}\ln L} = -y_{0}(L)y_{\phi}(L)
  \label{eq:RG}
\end{equation}
where $L$ is an infrared cutoff, and
\begin{equation}
    y_0 \propto K-1
\end{equation}
The flow diagram of is shown in figure 6,
from which we see that the SF-dimer transition is of the
Berezinnskii-Kostelitz-Thouless type, as is well known.
Since our model is isotropic, the renormalized value of $K$ should be
equal to unity, as already stated.
Then, when the system is in the SF state,
the starting point of the renormalization lies on the SF-N\'eel boundary
line which flows into the origin where $K=1$, and moves as ${\rm A \to B \to C}$,
as $\tilde J_2$ increases.
Finally the SF-dimer transition takes place when the starting point arrives at
the origin.
In the SF state, $y_\phi(L) = - y_0(L)$ in equation (\ref{eq:RG}),
resulting in
\begin{equation}
  y_{0}(L) = {y_0^{(0)} \over y_0^{(0)} \ln (L/L_0) + 1}
\label{eq:log}
\end{equation}
where $y_0^{(0)}$ is the bare value of $y_0(L)$ and $L_0$ is the cutoff length.
Then, there appear logarithmic corrections in various physical quantities
at every place in the SF region in our isotropic (i.e., $SU(2)$ symmetric) model.
On the SF-dimer critical point (origin O), on the other hand,
the logarithmic corrections vanish because $y_0 = y_\phi =0$.
This is very peculiar to the isotropic case.
Since the SF-dimer transition occurs at $y_\phi=0$,
one may think that the SF-dimer critical point can be obtained from
equations (\ref{eq:y-nnn}) or (\ref{eq:y-dd}).
However, as stated, the expressions (\ref{eq:y-nnn}) and (\ref{eq:y-dd}) are valid
only in the lowest order of $\Delta$, $\tilde J_{\rm NNN}$ and $\tilde J_3$,
although the critical properties are well expressed by the effective Hamiltonian.
Then it needs the numerical calculation for determining the SF-dimer critical
point even in case of $\tilde J_2=1$.

\begin{figure}[h]
   \begin{center}
      \scalebox{0.4}[0.4]{\includegraphics{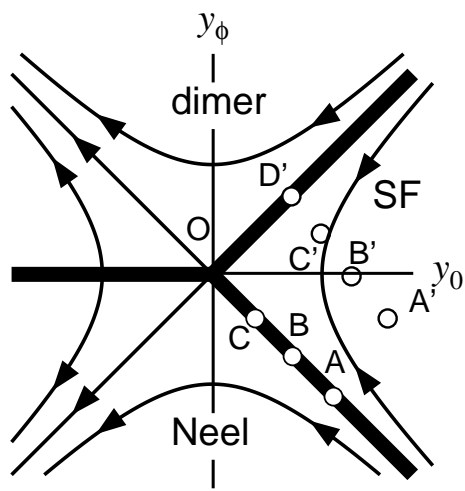}}
   \end{center}
   \caption{The renormalization flow of the effective
            Hamiltonian (\ref{eq:sg-nnn}).
            The phase boundaries are shown by thick lines.
            In our isotropic case, the starting point of the renormalization
            lies on the SF-N\'eel boundary line.
            As $\tilde J_2$ increases, the starting point moves
            as ${\rm A \to B \to C \to O}$. If the system has the $XXZ$ symmetry
            with $\Delta \equiv J^z/J^\perp < 1$,
            the starting point of the renormalization moves as
            ${\rm A' \to B' \to C' \to D'}$.}
\end{figure}

If the model has the $XXZ$ symmetry (no longer isotropic)
with $\Delta \equiv J^z/J^\perp < 1$,
the starting point of the renormalization moves as
${\rm A' \to B' \to C' \to D'}$ as $\tilde J_2$ increases.
When the starting point arrived at ${\rm D'}$,
the SF-dimer transition occurs.
Then,  in the $XXZ$ symmetric case,
the logarithmic corrections exist only at the SF-dimer critical point,
and do not exist in the SF region.

\bigskip
\section{Numerical result}

To confirm the consideration in \S2 and to obtain the ground-state
phase diagram on the $\tilde J_2 - \tilde J_3$ plane,
we performed the numerical diagonalization for finite
systems for $N=6,12,18,24$ by use of Lanczos algorithm under the periodic
boundary conditions.
It is very easy to distinguish whether the ground state is ferrimagnetic
($M=M_{\rm s}/3$) or $M=0$ from the numerical data.
However, it is difficult to detect the SF-dimer critical point from the
numerical data, because this transition is of
the Berezinskii-Kosterlitz-Thouless type \cite{Haldane,ON}
with pathological critical behavior.
Okamoto and Nomura (ON) \cite{ON} developed a method by use of which
the SF-dimer critical point of the $S=1/2$ NNN interaction model of figure 3
can be successfully determined from the numerical data for the energy gaps.
Let us explain this method , focusing on its physical meaning.
In usual cases, the ground state is unique (not twofold degenerate)
in finite systems, except for the special cases such as the Ising model and
the Majumdar-Ghosh model \cite{MG,Majumdar}.
How the twofold degenerate ground state is realized in infinite systems?
The energy gap of a low-lying excited state of finite systems
rapidly decreases as the system size $N$ increases,
and finally degenerate to the ground state in $N \to \infty$.
Then the linear combination of the ground state and the above-mentioned
excited state results in the twofold degenerate ground state of
the infinite systems.
In our case, the ground state of finite systems has the property
$S_{\rm tot}=0$ as far as it lies in the $M=0$ subspace
(i.e., except for the ferrimagnetic case).
The twofold degenerate dimer state of infinite systems also
has the property $S_{\rm tot}=0$.
Then the above-mentioned excited state should be also of $S_{\rm tot}=0$,
because of the law of the addition of the angular momentum.
Then we can conclude that the lowest excitation in finite systems
is of $S_{\rm tot}=0$ in the dimer region.
In the SF region, on the other hand, the lowest excitation should be
of the spin-wave type with $S_{\rm tot}^z=\pm 1$ (one magnon state).
In the present case, the excitation with the same energy exists
in the $S_{\rm tot}^z=0$ subspace due to the isotropic nature.
This means that the system has three-fold degenerate lowest excitation with
$S_{\rm tot}=1$, when it lies in the SF region.

From the above physical consideration, we can write down the criterion
\begin{eqnarray}
      \Delta E_{\rm ss}(N) &<& \Delta E_{\rm st}(N)
      ~~\Longleftrightarrow ~~\hbox{dimer state} \nonumber \\
      \Delta E_{\rm ss}(N) &>& \Delta E_{\rm st}(N)
      ~~\Longleftrightarrow ~~\hbox{spin fluid state}
  \label{eq:criterion}
\end{eqnarray}
where $\Delta E_{\rm ss}$(N) and $\Delta E_{\rm st}$(N) are the singlet-singlet
energy gap and singlet-triplet energy gap for a finite-size
system with $N$ spins, defined by
\begin{eqnarray}
    \Delta E_{\rm ss}(N) \equiv E_1(N,S^{\rm (tot)}=0) - E_{\rm g}(N) \\
    \Delta E_{\rm ss}(N) \equiv E_0(N,S^{\rm (tot)}=1) - E_{\rm g}(N)
\end{eqnarray}
respectively.
Here $E_0(N,S^{\rm (tot)})$ and $E_1(N,S^{\rm (tot)})$
are the lowest and second lowest
energies in the subspace with $S^{\rm (tot)}$ and
$E_{\rm g} = E_0(S^{\rm (tot)} =0)$, respectively.
This criterion can be obtained also by use of the effective Hamiltonian
representation, renormalization group method and the conformal field theory
\cite{ON,NO1,NO2}.

\begin{figure}[h]
   \begin{center}
      \scalebox{0.35}[0.35]{\includegraphics{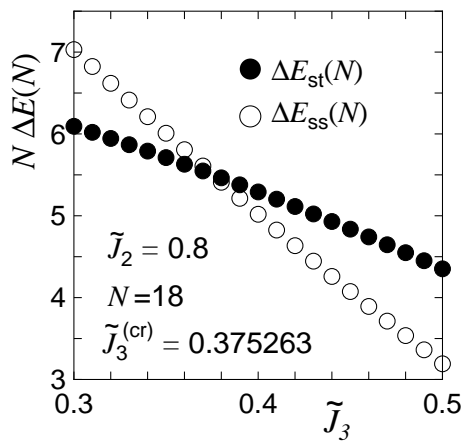}}
   \end{center}
   \caption{Level crossing between $\Delta E_{\rm ss}(N)$
            and $\Delta E_{\rm ss}(N)$
            when $\tilde J_2=0.8$ and $N=18$.
            From the crossing point, we obtain
            $\tilde J_3^{\rm (cr)}(N=18) = 0.375263$.}
\end{figure}

\begin{figure}[h]
   \begin{center}
      \scalebox{0.35}[0.35]{\includegraphics{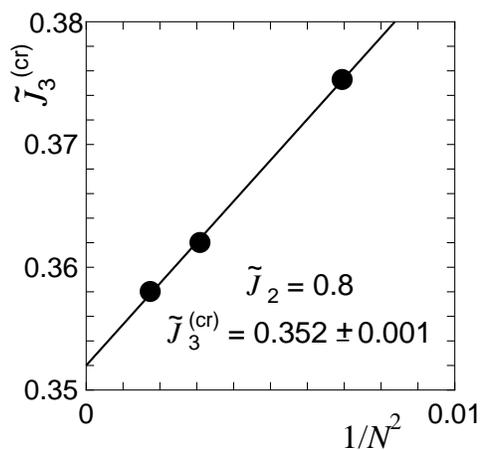}}
   \end{center}
   \caption{The extrapolation of $\tilde J_3^{\rm (cr)}$ to $N \to \infty$
            in $\tilde J_2 = 0.8$ case.
            From this, we see $\tilde J_3^{\rm (cr)} = 0.352 \pm 0.001$.}
\end{figure}

Figure 7 shows the crossing between $S_{\rm tot}=0$ and
$S_{\rm tot}=1$ excitations when $\tilde J_2=0.8$ and $N=18$.
By use of the interpolation, we see that the crossing point is
$\tilde J_3^{\rm (cr)}(N=18)=0.375263$.
We can obtain the SF-dimer critical point of the infinite system by extrapolating
$\tilde J_3^{\rm (cr)}(N)$ to $N \to \infty$.
The finite-size dependence of $\tilde J_3^{\rm (cr)}(N)$ has the form
\begin{equation}
    \tilde J_3^{\rm (cr)}(N)
    = \tilde J_3^{\rm (cr)}(\infty) + ({\rm const}/N^2)
\end{equation}
due to the existence of the irrelevant fields, as was discussed in \cite{ON,NO1,NO2}.
Figure 8 shows the extrapolation of $\tilde J_3^{\rm (cr)}(N)$ to
$N \to \infty$ in case of $\tilde J_2=0.8$,
resulting in $\tilde J_3^{\rm cr} = 0.352 \pm 0.001$.

By sweeping parameters, we finally obtain the phase diagram on the
$\tilde J_2 - \tilde J_3$ plane.
The result is shown in figure 9.
We note that the point $(\tilde J_2, \tilde J_3)$ is equivalent to
the point $(\tilde J_2/\tilde J_3, 1/\tilde J_3)$
by interchanging the role of $J_1$ and $J_3$,
as already stated in \S1.

\begin{figure}[h]
   \begin{center}
         \scalebox{0.35}[0.35]{\includegraphics{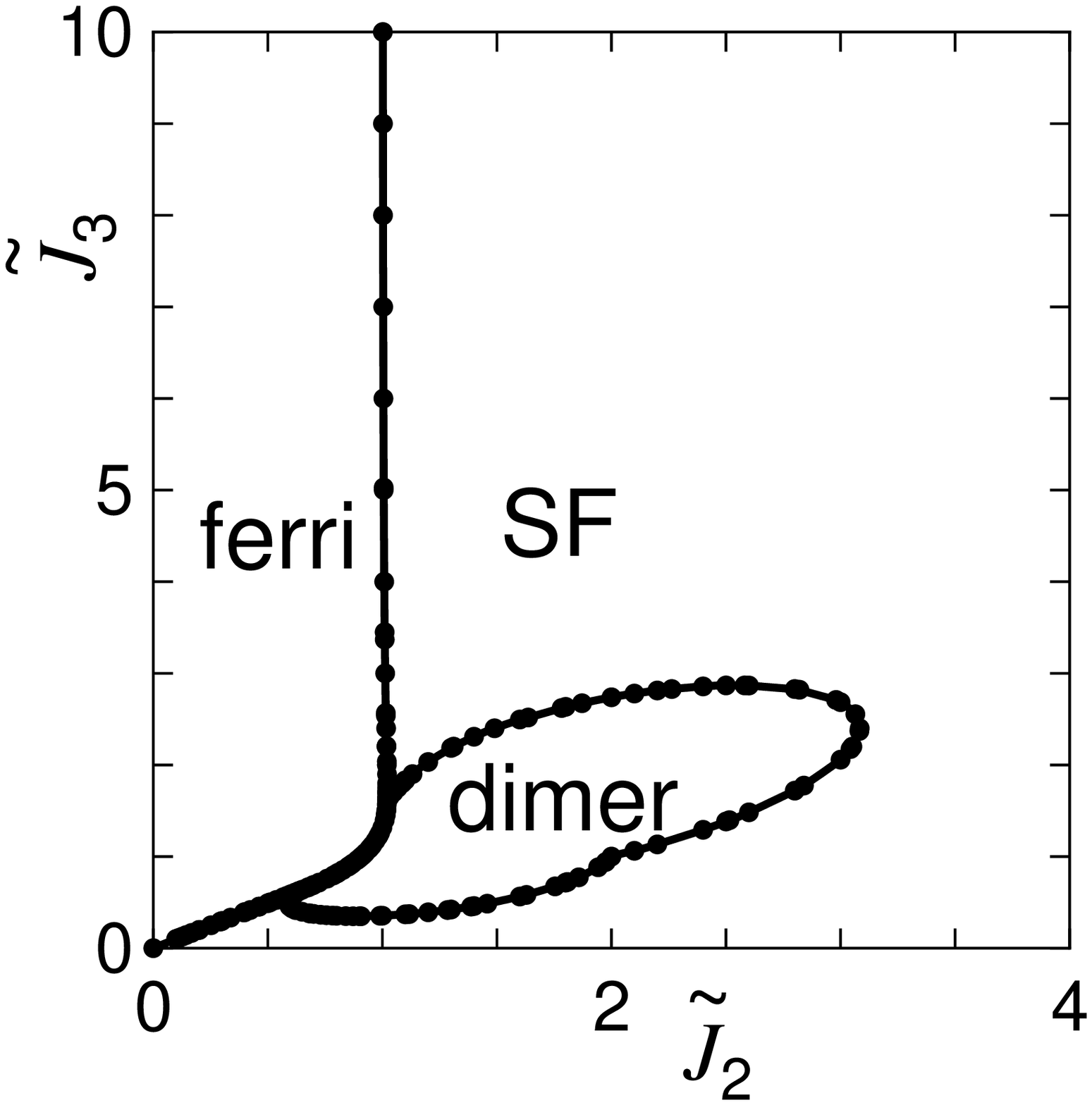}}
         \scalebox{0.35}[0.35]{\includegraphics{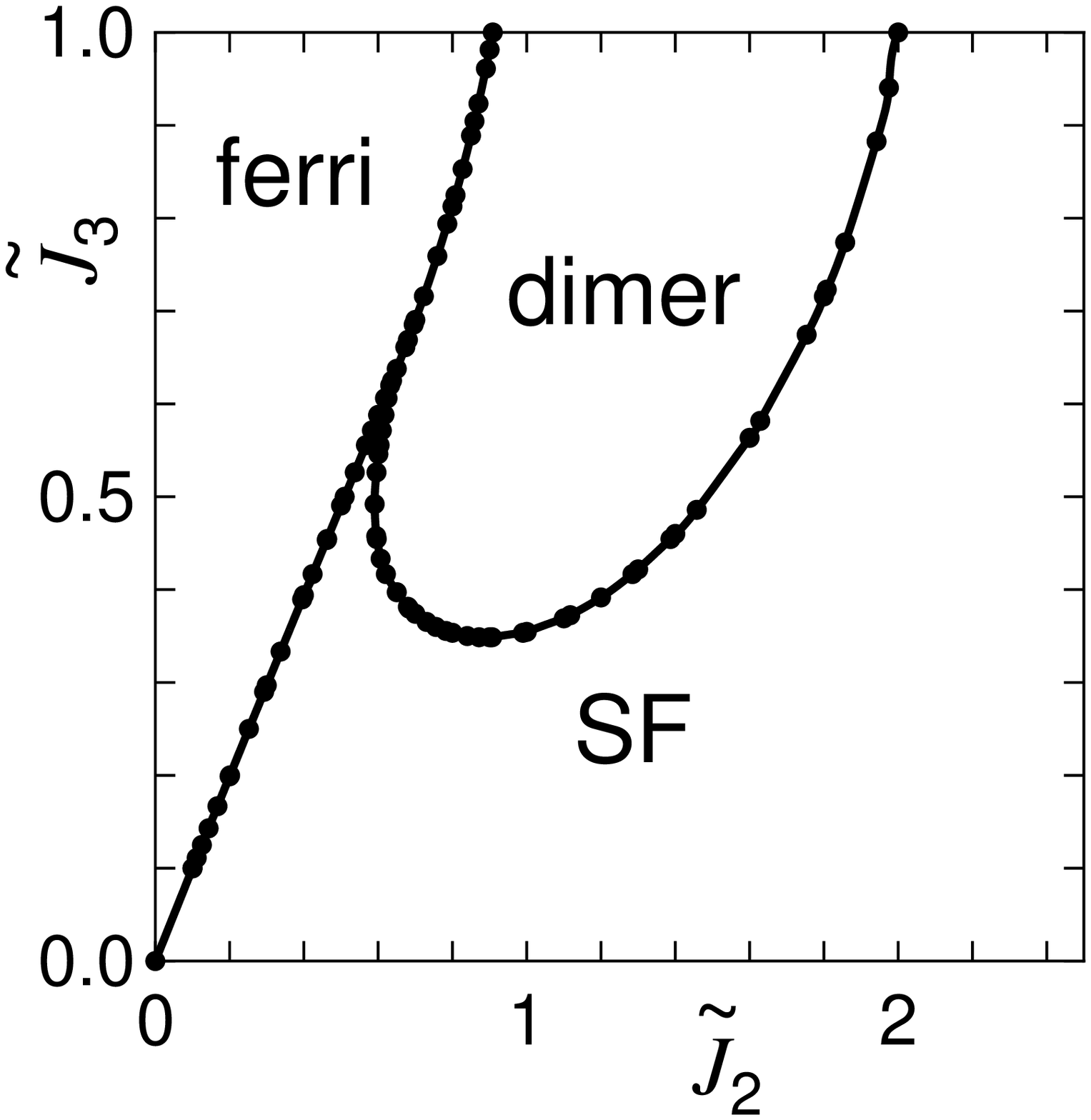}}
   \end{center}
   \caption{Phase diagram of the DD chain model.
            The estimated errors in the critical values
            are less than 0.001.
            The $\tilde J_3=1$ case is reduced to the model of
            Takano et al. (see \S4).}
\end{figure}

\bigskip
\section{Discussion}

In \S2, we have stated that the ground-state quantum phase transition of the
DD chain model (the present model) has the same universality class as that of the
NNN model (figure 3).
We have confirmed this analytically by use of the effective Hamiltonian
representation.
Here we also confirm this by numerical method.

\begin{figure}[h]
   \begin{center}
      \scalebox{0.5}[0.5]{\includegraphics{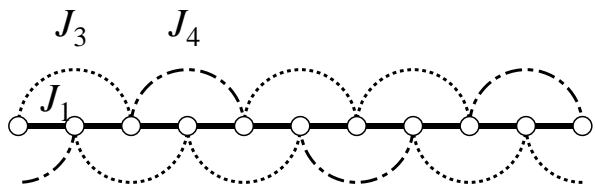}}
   \end{center}
   \caption{The interpolation model between the DD chain model
            with $\tilde J_2=1$ and the NNN interaction model.
            Solid lines denote $J_1$, dotted lines $J_3$ and dot-dashed lines
            $J_4$.}
\end{figure}

Let us consider the model of figure 10
which interpolates between the DD chain model with $\tilde J_2=1$ (figure 2)
and the NNN interaction model (figure 3).
When $J_4/J_3=0$ the interpolation model is reduced to the DD chain model with
$\tilde J_2=1$, and when $J_4/J_3=1$ to the NNN interaction model.
Figure 11 shows the level crossings between $\Delta_{\rm st}(N)$
and $\Delta_{\rm ss}(N)$ for $J_4/J_3=0$ and $J_4/J_3=1$ cases.
The behavior of level crossing is essentially the same
when the parameter $J_4/J_3$ runs from $J_4/J_3=0$ to $J_4/J_3=1$,
as can be seen from figure 11.
Furthermore any other excitations cross them between $J_4/J_3=0$ to $J_4/J_3=1$.
Figure 12 shows the SF-dimer critical point $J_3^{\rm cr}$ of the interpolation model,
in which the critical point smoothly changes.
Thus we can safely conclude that the ground-state quantum phase transition of the
DD chain model (the present model) has the same universality class as that of the
NNN interaction model (figure 3).
The DD chain model with $\tilde J_2=1$ is obtained from the NNN interaction 
model by removing
one NNN interaction in every three NNN interactions, as already stated.
Instead of removing, a similar (but not exact) effect may be realized
by decreasing the strength of the NNN interaction to $2/3$ of
the original strength.
If this is the case, the SF-dimer critical point of the DD chain model with $\tilde J_2=1$
is $3/2$ of that of the NNN interaction model, which results in
$\tilde J_3^{\rm (cr)} = (3/2)\times 0.2411 \simeq 0.36$.
This semiqualitatively explains our numerical result
$\tilde J_3^{\rm (cr)}=0.354 \pm 0.001$ when $\tilde J_2=1$.
This fact also appears in equations (\ref{eq:y-nnn}) and (\ref{eq:y-dd}).

\begin{figure}[h]
   \begin{center}
      \scalebox{0.35}[0.35]{\includegraphics{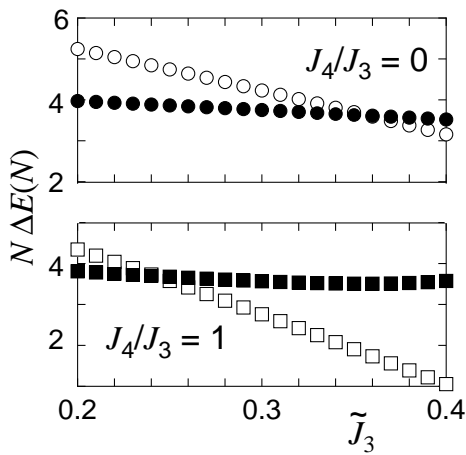}}
   \end{center}
   \caption{Level crossing of the interpolation model when $J_4/J_3 = 0$
            and $J_4/J_3 = 1$ for $N=18$.
            Closed circles and squares represent $\Delta_{\rm st}(N)$,
            and open circles and squares $\Delta_{\rm ss}(N)$.}
\end{figure}

\begin{figure}[h]
   \begin{center}
      \scalebox{0.35}[0.35]{\includegraphics{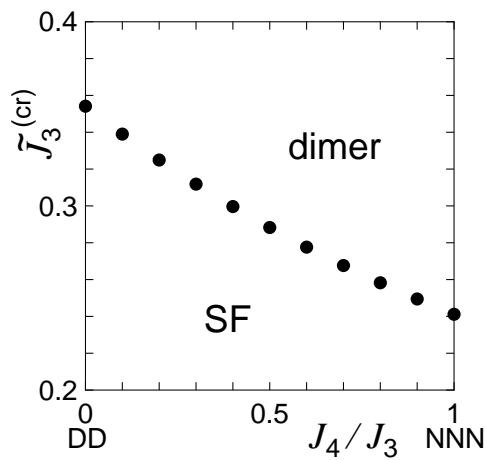}}
   \end{center}
   \caption{The SF-dimer critical points of the interpolation model
            between the DD chain model with $\tilde J_2=1$ 
            and the NNN interaction model.}
\end{figure}

Takano, Kubo and Sakamoto (TKS) \cite{Takano} investigated the $J_3 = J_1$
case of the present DD chain model (see figure 13(a)).
They concluded that the ground state of their model is composed of
three phases.
The ferrimagnetic phase ($M=M_{\rm s}/3$) appears when $\tilde J_2 < 0.909$.
In the tetramer-dimer (TD) phase, which appears when $0.909 < \tilde J_2 <2$,
the state is exactly the regular array of
tetramers and dimers as shown in figure 13(b).
Figure 13(c) shows the dimer-monomer (DM) state,
appearing when $\tilde J_2>2$, which is composed
of the regular array of the singlet dimers and free spins.
Because of the free spins, the DM state is macroscopically degenerate.

Let us discuss the relation between our model and TKS's model.
In the DM state of TKS's model, the monomers are completely free
but this is very peculiar to this model.
In our model, since the symmetry of a diamond is broken because $\tilde J_3 \ne 1$,
the monomer is no longer free and has an effective interaction between
neighboring monomers through the dimer between them.
Therefore the DM state of TKS's model is smoothly connected to the spin-fluid state
of our model, as can be seen in figure 9.
The tetramer in the TD state is also special to TKS's model.
When the symmetry of the diamond is broken,
the tetramer is decomposed into two dimers existing on stronger bonds,
as is shown in figure 5.
Then the TD state of TKS's model is a special case of the dimer state
of our DD chain model.
Thus the physical pictures of our model and TKS's model are
consistent with each other.

\begin{figure}[h]
   \begin{center}
       \hbox{~~~~~~~~~~~~~~~~~\Large (a)}
       \scalebox{0.5}[0.5]{\includegraphics{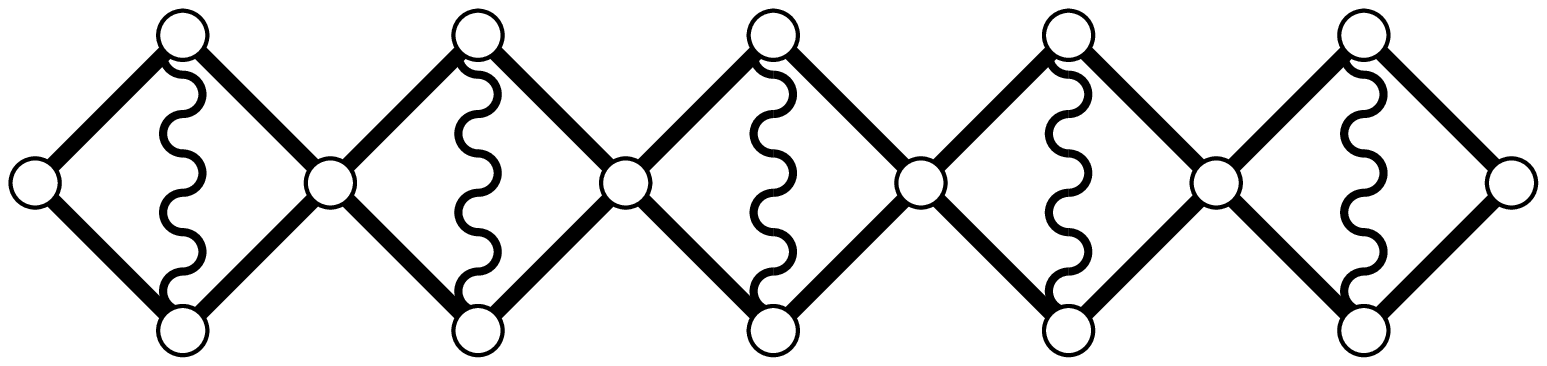}}
    \end{center}
   \begin{center}
       \hbox{~~~~~~~~~~~~~~~~~\Large (b)}
       \scalebox{0.5}[0.5]{\includegraphics{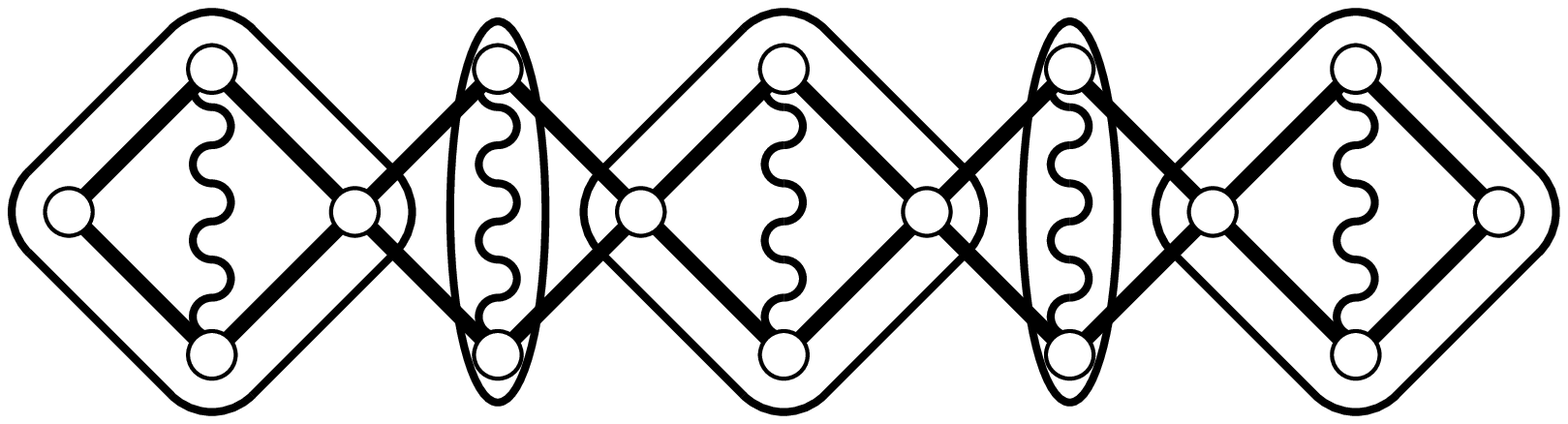}}
   \end{center}
   \begin{center}
       \hbox{~~~~~~~~~~~~~~~~~\Large (c)}
       \scalebox{0.5}[0.5]{\includegraphics{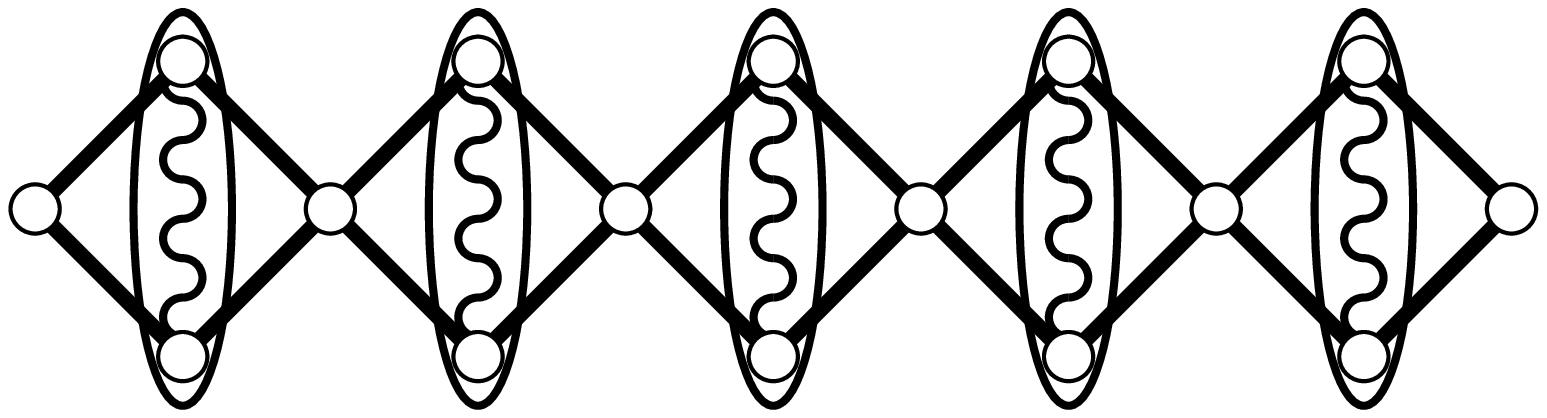}}
   \end{center}
   \caption{(a) The model of Takano, Kubo and Sakamoto.
            (b) The tetramer-dimer (TD) state. The rectangles represent tetramers
            and the ellipses singlet dimers.
            (c) The dimer-monomer (DM) state.}
\end{figure}

We can confirm the reliability of our numerical results by checking the
properties of excitations.
At the SF-dimer critical point, the cos-term of the effective Hamiltonian
(\ref{eq:sg-nnn}) vanishes and the relation
$\Delta E_{\rm ss}(N) = \Delta E_{\rm st}(N)$ holds,
as discussed by Okamoto and Nomura \cite{ON,NO1,NO2}.
Then the system is purely Gaussian and has the low-lying excitation
energies proportional to $1/N$ in finite systems.
The lowest order correction to $1/N$ may be of $1/N^3$.
This correction comes from
the band curvature and the wave-number dependence of the coupling constant
of the interaction between Jordan-Wigner fermions, which was neglected
in the course of deriving the effective Hamiltonian.
Figure 14 shows the $1/N^2$-dependence of 
$N \Delta E(N)$ at $\tilde J_2=1.2$ and $\tilde J_3=0.39$,
which is the SF-dimer critical point
where $\Delta E_{\rm ss}(\infty) = \Delta E_{\rm st}(\infty)$.
As can be seen from figure 14, the size dependence of the
lowest excitation is well expressed as
\begin{equation}
    N\Delta E(N) = a + (b/N^2)
\end{equation}
which is consistent with the above-mentioned discussion.
The quantity $a$ is related to the spin wave velocity as
\begin{equation}
    a = 2\pi v_{\rm s} x
\end{equation}
where $x$ is the scaling dimension of this excitation, which is equal to $1/2$
at the critical point \cite{ON,NO1,NO2}.
Since $a=3.851 \pm 0.001$ in case of figure 14,
we obtain
\begin{equation}
    v_{\rm s} = 1.226 \pm 0.001
    \label{eq:vs-exc}
\end{equation}

\begin{figure}[h]
   \begin{center}
      \scalebox{0.35}[0.35]{\includegraphics{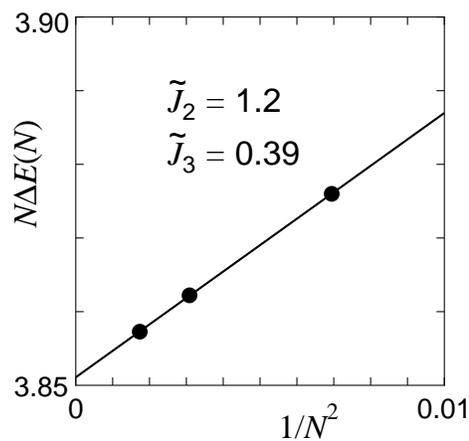}}
   \end{center}
   \caption{System-size dependence of the scaled excitation gap $N\Delta E$
            on the SF-dimer critical point.}
\end{figure}

The system-size dependence of the ground-state energy also provides us with
useful information.
Under periodic boundary conditions, it is written as \cite{BCN,Affleck}
\begin{equation}
    {E_{\rm g}(N) \over N}
    = \epsilon_{\rm g}(\infty) - {\pi v_{\rm s}c \over 6N^2} + \cdots
    \label{eq:gs-energy}
\end{equation}
where $E_{\rm g}(N)$ is the ground-state energy of the $N$-spin systems,
$\epsilon_{\rm g}(\infty)$ the ground state energy of the infinite system
per spin, $v_{\rm s}$ the spin wave velocity, and $c$ the conformal charge
which is equal to unity in our universality class.
Figure 15 shows the system-size dependence of
the ground state energy in case of $\tilde J_2=1.2$ and $\tilde J_3=0.39$.
From the slope of the line, we obtain
\begin{equation}
    v_{\rm s} = 1.23 \pm 0.01
    \label{eq:vs-gs}
\end{equation}
which well agrees with equation (\ref{eq:vs-exc}).
The fact that the values of $v_{\rm s}$ given by equations (\ref{eq:vs-exc})
and (\ref{eq:vs-gs}) agree within the numerical error justifies our numerical
analysis.

\begin{figure}[h]
   \begin{center}
      \scalebox{0.35}[0.35]{\includegraphics{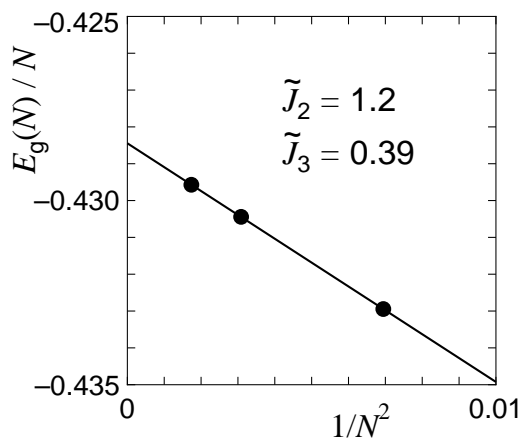}}
   \end{center}
   \caption{System-size dependence of the ground-state energy $E_{\rm g}(N)$
            on the SF-dimer critical point.}
\end{figure}

The spin wave velocity $v_{\rm s}$ can be also obtained from the
lowest excitation having $S_{\rm tot}^z=0$ and $k=2\pi/N$ by
\begin{equation}
    v_{\rm s}
    = \lim_{N \to \infty}{N \Delta E (N, S_{\rm tot}^z=0,k=2\pi/N) \over 2\pi}
    \label{eq:vs-formula}
\end{equation}
We note that this formula is free from the logarithmic corrections
even in the SF region \cite{NO2}.
Figure 16 shows this extrapolation procedure, which brings about
$v_{\rm s}=1.228 \pm 0.001$.
This is also consistent with equation (\ref{eq:vs-exc}) and equation (\ref{eq:vs-gs}).

\begin{figure}[h]
   \begin{center}
      \scalebox{0.35}[0.35]{\includegraphics{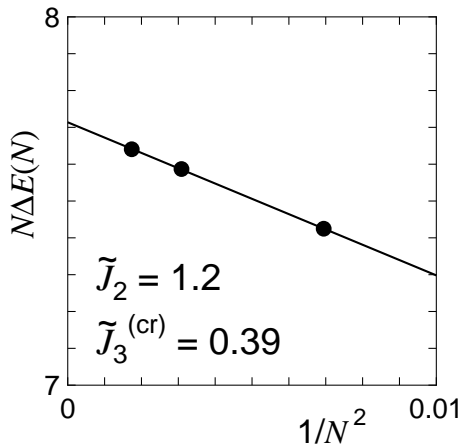}}
   \end{center}
   \caption{Extrapolation procedure of equation (\ref{eq:vs-formula}).
            From the intersection, we obtain $v_{\rm s}=1.228 \pm 0.001$.}
\end{figure}

Let us discuss the logarithmic corrections in the SF region.
As stated in \S2, there appear logarithmic corrections in various physical
quantities when the system is in the SF region.
This is very peculiar to our isotropic case.
In the following we check this point numerically.
The singlet-singlet gap and the singlet-triplet gap are expressed as
\begin{equation}
    \Delta E_{\rm ss}(N) = {2\pi v_{\rm s} x_{\rm ss} \over N}~~~~~~
    \Delta E_{\rm st}(N) = {2\pi v_{\rm s} x_{\rm st} \over N}
    \label{eq:ss-st}
\end{equation}
respectively, where $x_{\rm ss}$ and $x_{\rm st}$ are the scaling dimensions
\begin{equation}
    x_{\rm ss} = {1 \over 2} \left( 1 + {3 \over 2}y_0(N) \right)
    ~~~~~~~
    x_{\rm st} = {1 \over 2} \left( 1 - {1 \over 2}y_0(N) \right)
    \label{eq:x-ss-st}
\end{equation}
with $y_0(N)$ given in equation (\ref{eq:log}).
It is difficult to directly detect the logarithmic dependence in 
equations (\ref{eq:ss-st})
and (\ref{eq:x-ss-st}) from the numerical data for $\Delta E_{\rm ss}(N)$
and $\Delta E_{\rm st}(N)$.
Because the logarithmic corrections are very slowly varying with respect to
the system size $N$,
its effects are actually observed as the change in
the spin wave velocity $v_{\rm s}$ between
$\Delta E_{\rm ss}$ and $\Delta E_{\rm st}$.
As an example, let us take the $\tilde J_2=1.2$ and $\tilde J_3=0.35$ point 
which lies in the SF region.
In fact, as shown in figure 17,
the spin wave velocities are estimated to be $v_{\rm s}=1.362 \pm 0.001$
and $v_{\rm s}=1.248 \pm 0.001$
from $\Delta E_{\rm ss}$ and $\Delta E_{\rm st}$, respectively.
Okamoto and Nomura \cite{ON,NO1,NO2} used the \lq\lq averaged excitation\rq\rq
\begin{equation}
    \Delta E_{\rm ave}(N)
    = {1 \over 4} \left\{ \Delta E_{\rm ss}(N) + 3\Delta E_{\rm st}(N) \right\}
\end{equation}
in which the lowest order logarithmic corrections vanish,
as can be seen from equation (\ref{eq:x-ss-st}).
From figure 17 we obtain $v_{\rm s}=1.277 \pm 0.001$ 
by use of $\Delta E_{\rm ave}$.
When we calculate the spin wave velocity through the formula (\ref{eq:vs-formula}),
we obtain $v_{\rm s}=1.277 \pm 0.001$, which shows very good agreement with that from
$\Delta E_{\rm ave}$.
We note that there is no logarithmic correction in  formula (\ref{eq:vs-formula}),
as already stated.
Thus our numerical analysis is consistent with our consideration in \S2
with respect to the logarithmic corrections,
although we could not directly observe its existence.
It may need systems with several thousand (or more) spins to directly observe
the contribution of the logarithmic corrections.

\begin{figure}[h]
   \begin{center}
      \scalebox{0.35}[0.35]{\includegraphics{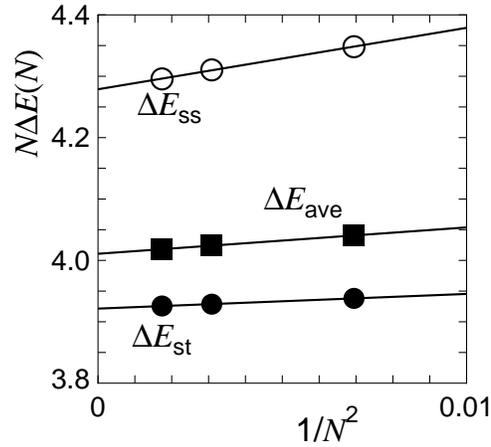}}
   \end{center}
   \caption{Behavior of $\Delta E_{\rm ss}$, $\Delta E_{\rm st}$ and
            $\Delta E_{\rm ave}$ at the point 
            $(\tilde J_2,\tilde J_3)=(1.2,0.35)$ in the SF region.
            From the intersection of the $\Delta E_{\rm ave}$ line,
            we obtain $v_{\rm s}=1.277 \pm 0.001$.
            The apparent spin wave velocities are
            $v_{\rm s}=1.362 \pm 0.001$ from $\Delta E_{\rm ss}$ and
            $v_{\rm s}=1.248 \pm 0.001$ from $\Delta E_{\rm st}$.}
\end{figure}

In summary, we have explained that the frustration brings about the non-magnetic
ground state in our DD chain model by use of the analytical method,
physical consideration and numerical method.
We have also obtained the phase diagram on the
$\tilde J_2 - \tilde J_3$ plane numerically.

\bigskip
\section*{Acknowledgement}

We would like to express our appreciation to H. Tanaka for stimulating
discussion and also for showing us the experimental result by his group
prior to publication.
A part of the numerical calculation was done by use of program package
TITPACK Ver.2 coded by H. Nishimori.

\end{document}